Electroresistance effects in ferroelectric tunnel barriers

D. Pantel[*] and M. Alexe

Max-Planck-Institute of Microstructure Physics, Weinberg 2, 06120 Halle, Germany

**Abstract**

Electron transport through fully depleted ferroelectric tunnel barriers sandwiched between two metal electrodes and its dependence on ferroelectric polarization direction are investigated. The model assumes a polarization direction dependent ferroelectric barrier. The transport mechanisms, including direct tunneling, Fowler-Nordheim tunneling and thermionic injection, are considered in the calculation of the electroresistance as a function of ferroelectric barrier properties, given by the properties of the ferroelectric, the barrier thickness, and the metal properties, and in turn of the polarization direction. Large electroresistance is favored in thicker films for all three transport mechanisms but on the expense of current density. However, switching between two transport mechanisms, i.e., direct tunneling and Fowler-Nordheim tunneling, by polarization switching yields a large electroresistance. Furthermore, the most versatile playground in optimizing the device performance was found to be the electrode properties, especially screening length and band offset with the ferroelectric.

---

[*] Corresponding author, dpantel@mpi-halle.mpg.de

## I. INTRODUCTION

In recent years, the preparation of high quality ultra thin films of perovskite ferroelectrics (F) by various deposition techniques such as pulsed laser deposition or off-axis sputtering has pushed the lower limit of the critical thickness for ferroelectricity to a few unit cells,[1,2,3,4] in good agreement with theory.[5] This was the prerequisite for observing direct quantum-mechanical tunneling through a ferroelectric barrier[6,7] for which giant electroresistance (ER) was predicted.[8] A direct tunneling (DT) effect and a resistance state depending on the polarization direction were experimentally shown on tunneling junctions with a ferroelectric barrier.[9, 10, 11, 12] However, depending on the effective thickness of the ferroelectric barrier other transport mechanisms such as Fowler-Nordheim tunneling (FNT) or thermionic injection (TI) rather than the direct tunneling may play the major role in transport across ultrathin ferroelectric films, as has been already shown on only 5nm thick $BiFeO_3$ thin films.[13] The various transport mechanisms which concurrently contribute to the effective carrier conduction in such a metal-ferroelectric-metal (MFM) structure raise the questions which mechanism governs the electronic transport and what are the parameters to be tuned in order to maximize the resistance change, or the ER.

In the present paper we analyze the transport through a MFM structure, as shown in Fig. 1(a). A perovskite ferroelectric, such as $BaTiO_3$ (BTO), acts as a potential barrier. The effective barrier is polarization direction dependent due to insufficient screening of the polarization charges.[8,7] In our analysis we included three major contributions to the effective current, namely direct tunneling, Fowler-Nordheim tunneling, and thermionic injection.

## II. MODEL

We study here a metal-ferroelectric-metal heterostructure as sketched in Fig. 1(a). The F with a spontaneous polarization $P_S$ and a static permittivity $\varepsilon_{stat}$ is sandwiched between two different metals ($M_i$) with different Thomas-Fermi screening lengths $l_i$. A voltage is applied to metal $M_2$ and the resulting current $j$ through the MFM heterostructure is calculated in dependence on the ferroelectric polarization direction.

We assume that the ferroelectric layer is fully depleted, i.e., there are no free carriers within the ferroelectric, and the semiconductor properties of the ferroelectric do not play any role in the transport mechanism. This assumption is reasonable up to a certain film thickness $d \approx 2w$, when $w$ is the depletion width, which depends on the effective free carrier concentration $N_{eff}$, the apparent build-in potential $V_{bi}$', which is itself polarization dependent,[14] and the static permittivity $\varepsilon_{stat}$, and can be calculated by[15]

$$w = \sqrt{\frac{2\varepsilon_0 \varepsilon_{stat}(V + V_{bi}^{'})}{eN_{eff}}} \tag{1}$$

For $d \approx 2w$ at zero applied voltage $V$ the film is fully depleted. For instance with $N_{eff} = 5\times10^{19}$ 1/cm$^3$, $V_{bi}$' = 0.1V, and $\varepsilon_{stat}$ = 60 the depletion width $w$ is about 3.5nm, thus we restrict our investigation to film thicknesses $d$ smaller than about 5nm.

The polarization $P$ is perpendicular to the film surface, either pointing toward the contact where the voltage is applied ($P > 0$) after switching with a negative voltage or away from it ($P < 0$) after switching with a positive voltage as shown in Fig. 1(a)). This resembles the polarization state which is typically found in single domain tetragonal thin ferroelectric films such as BTO (Ref. 16) or PbTiO$_3$.[17]

We assume also that the currents are mostly electron currents because in most oxide ferroelectrics with metal electrodes the barrier for electrons is lower than the barrier for holes.[18] It is noteworthy, that this applies for *n*-type as well as for *p*-type ferroelectric films.[19]

The carrier transport through the MFM structure depends on the barrier properties, which in turn depend on the polarization. The polarization direction may influence the electrostatic potential of the barrier because of the polarization charges or the barrier's thickness due to piezoelectric strains.[7] Depending on the piezoelectric coefficient of the ferroelectric material the influence of the piezoelectric strains might be neglected, as it is the case for La$_{0.1}$Bi$_{0.9}$MnO$_3$ tunneling barriers[9] or may play a crucial role for the transport through the barrier.

However, here, we focus on the polarization dependence of the electrostatic potential in the MFM heterostructure and introduce it by employing a model proposed by Zhuravlev *et al.*.[8] The ferroelectric polarization $P$ creates a surface charge equal to $P$ at the metal-ferroelectric interfaces [see plus and minus signs in Fig. 1(a)]. These surface charges are screened inside the metal as sketched in Fig. 1(a) by the shaded areas, but due to the finite capacitance of the screening space charges of real metals the screening is incomplete inside the ferroelectric. This leads to a depolarization field opposing the polarization direction.[20] Furthermore, the asymmetry induced by the polarization direction and the different screening abilities of the metals gives different shapes of the potential barrier for different polarization directions.[8]

The imperfect screening originates from the finite capacitance of the screening space charges[21,22] which is due to the small but finite screening length $l_{1,2}$ in the metal electrodes $M_{1,2}$ and the ionic permittivity of the electrode material $\varepsilon_M$.[23] The screening lengths can be calculated by the free electron model and are on the order of 1Å [e.g. for Cu: 0.55Å (Ref. 24) and for SrRuO$_3$ (SRO): 0.8Å (Ref. 23)]. Only few reports give ionic permittivity $\varepsilon_M$ of the electrode materials, but for SRO $\varepsilon_M$ is found to be about 8.[23] The change in the potential barriers $2\Delta\Phi_{1,2}$ on polarization reversal and the depolarization field $E_{depol}$ inside the ferroelectric can be calculated from Thomas-Fermi screening (see, e.g., Ref. 25) bearing in mind that the polarization charge is a two-dimensional sheet of immobile charges at the interface between ferroelectric and metal as sketched in Fig. 1(a),[8,20]

$$\Delta\Phi_i = \frac{l_i Q_s}{\varepsilon_0 \varepsilon_{M,i}} e, \quad i = 1, 2, \tag{2}$$

$$E_{depol} = -\frac{(P - Q_s)}{\varepsilon_0 \varepsilon_{stat}}, \tag{3}$$

with the screening charge density $Q_s$, given by

$$Q_s = \frac{Pd}{\varepsilon_{stat}\left(\frac{l_1}{\varepsilon_{M,1}} + \frac{l_2}{\varepsilon_{M,2}}\right) + d}. \tag{4}$$

Here, the initial assumption of a fully depleted ferroelectric film is used. Note that here $P$ is the polarization charge which contributes to the depolarization field, which is not necessarily the spontaneous polarization as discussed below.

To arrive at the potential barrier relevant for transport processes the potential due to band offsets and the image force potential have to be superimposed. If the permittivity responsible for image force lowering is sufficiently high, the image force lowering can be neglected as a first approximation. The potential barrier without image force lowering is then given by

$$\Phi_{B,i} = \Phi_i \pm \Delta\Phi_i, \quad (5)$$

where $\Phi_{1,2}$ is the barrier without polarization which might be different for different metals.[15,18] The upper sign (+) applies for $\Phi_1$ (the barrier at the interfaces between metal $M_1$ and the ferroelectric) and the lower sign (-) for $\Phi_2$. The resulting potential-energy profile for $P > 0$ is shown in Fig. 1(b) as a continuous line and for $P < 0$ as a dashed one.

Although here the depolarization field is assumed to be caused by the insufficient screening, a thin passive layer between ferroelectric and electrode would lead qualitatively to the same effects. The same applies for other screening mechanisms which might be different from Thomas-Fermi screening.[26]

Substantially different to the above-mentioned screening mechanism is the ionic screening.[22, 27] In this scenario, part of the polarization charges penetrate into the electrode, which is feasible by the ionic structure of oxide electrodes, and is then screened in situ.[22] That means the screening charges and polarization charges are at the same place yielding no depolarization field. If one part of the polarization is screened by ionic screening and the other by Thomas-Fermi screening, the polarization contributing to the depolarization field, and therefore entering Eq. (4), is reduced.

The potential barrier and the depolarization field are subsequently used to calculate the current $j$ across the heterostructure when a voltage $V$ is applied to metal $M_2$. Since $j$ depends on the barrier and hence on the polarization direction we can define the electroresistance *ER*, given by the "pessimistic" definition

$$ER = \frac{j(P>0) - j(P<0)}{j_>}, \quad (6)$$

where $j_>$ denotes the current density in the high conduction state.

In our calculations we took account of direct tunneling and the interface-limited mechanisms thermionic injection and Fowler-Nordheim tunneling. We did not include bulk-limited conduction mechanisms, like space-charge-limited current, into this study because they are not expected to influence the transport through ultrathin films.

A. Direct tunneling

Direct tunneling is a quantum-mechanical phenomenon. For the present analysis we used the current density $j_{DT}$ given by Gruverman et al. [12] for a trapezoidal potential barrier [see Fig. 1(b)] using the WKB approximation,

$$j_{DT} = C \frac{\exp\left[\alpha\left\{\left(\Phi_{B,2} - \frac{eV}{2}\right)^{3/2} - \left(\Phi_{B,1} + \frac{eV}{2}\right)^{3/2}\right\}\right]}{\alpha^2 \left[\sqrt{\Phi_{B,2} - \frac{eV}{2}} - \sqrt{\Phi_{B,1} + \frac{eV}{2}}\right]^2} \sinh\left[\frac{3eV}{4}\alpha\left\{\sqrt{\Phi_{B,2} - \frac{eV}{2}} - \sqrt{\Phi_{B,1} + \frac{eV}{2}}\right\}\right]$$

(7)

where $C = -\dfrac{4em_{e,ox}}{9\pi^2\hbar^3}$, $\alpha = \dfrac{4d\sqrt{2m_{e,ox}}}{3\hbar(\Phi_{B,1} + eV - \Phi_{B,2})}$, and $m_{e,ox}$ being the effective tunneling electron mass. Because a finite image force lowering is not included in the trapezoidal barrier, the current given by the above formula rather underestimates the tunneling current.[28]

B. Thermionic injection

Thermionic injection describes the current which is due to charge carriers which overcome the potential barrier by thermal energy.[15] The barrier height is lowered by image force lowering, called the Schottky effect. The current density can be described for sufficiently high voltages (approx. $V > 100$mV at roomtemperature, i.e. approximately $3k_BT/e$) by

$$j_{Schottky} = A^{**}T^2 \exp\left[-\dfrac{1}{k_BT}\left(\Phi_B - \sqrt{\dfrac{e^3E}{4\pi\varepsilon_0\varepsilon_{ifl}}}\right)\right], \qquad (8)$$

where $\Phi_B$ is the potential barrier, $A^{**}$ the effective Richardson's constant, and $\varepsilon_{ifl}$ the permittivity of the ferroelectric responsible for image force lowering. At low voltages where the above formula is not valid we approximate the current by an ohmic relation.

C. Fowler-Nordheim tunneling

FNT is tunneling across a triangular-shaped potential barrier, which is formed by applying an electrical field $E$ to a rectangular or trapezoidal barrier.[29] FNT is basically the same physical phenomena as direct tunneling, but in a different voltage regime, i.e., the high-voltage regime. The current density is given by[29]

$$j_{FN} = \frac{e^3 m_e}{8\pi h m_{e,ox} \Phi_B} E^2 \exp\left[-\frac{8\pi\sqrt{2m_{e,ox}}}{3he} \frac{\Phi_B^{3/2}}{E}\right]. \qquad (9)$$

While image force lowering is essential for thermionic injection it will not seriously alter the FNT current at room-temperature or below, i.e., much below the Fermi temperature.[29]

In both the latter mechanisms the electric field $E$ is the field responsible for band tilting, i.e., a superposition of the applied field $E_{ap} = -V/d$, the depolarization field $E_{depol}$ and the field due to band alignment $E_{Band} = \frac{\Phi_2 - \Phi_1}{ed}$. The potential barrier $\Phi_B$ is the energy barrier which the electrons must overcome during transport across the MFM heterostructure, i.e., $\Phi_{B,1}$ for $V > 0$ and $\Phi_{B,2}$ for $V < 0$.

For our calculations which were performed using Wolfram Mathematica 6 we assume a MFM heterostructure at room temperature ($T = 300$K) with a perovskite bottom electrode $M_1$ made of SRO, BTO as the ferroelectric and Cu as the top electrode $M_2$ material. The parameters corresponding to BTO ($d = 3.2$nm, $P = 3\mu$C/cm$^2$,[30] $\varepsilon_{stat} = 60$, $\varepsilon_{ifl} = 10$, $m_{e,ox} = m_e$, and $A^{**} = 10^6$Am$^{-2}$K$^{-2}$) and its interface with SRO ($l_1 = 0.8$Å, $\Phi_1 = 1$V, $\varepsilon_{M,1} = 8$) and Cu [$l_2 = 0.55$Å, $\Phi_2 = 1$V, and $\varepsilon_{M,2} = 2$ (Ref. 31)] were used in the following unless other parameters are mentioned. We discuss and justify these values in Sec. IV.

**III. RESULTS**

Figure 2 shows the current densities resulting from the three conduction mechanisms versus applied voltage $V$ for the two opposite polarization states ($P > 0$ and $P < 0$ with solid and dashed lines, respectively). Direct tunneling gives rather parallel current branches for the two polarization directions, with the current for $P > 0$ (for $P > 0$ the polarization points towards the electrode with the larger ratio of screening length and permittivity $l_i/\varepsilon_M$, i.e., toward the electrode with lower screening ability) being larger than the one for $P < 0$ for all $d$ and $V$. Therefore the corresponding ER is positive and rather independent of the applied voltage. The current-voltage curve for FNT increases more steeply as the bias increases and the two polarization branches cross each other at a certain voltage. The ER, for negative voltages, is basically -1 at low $V$, goes through zero at a certain $V$ (here, approximately 1V) and then increases to positive values. Symmetrical behavior but opposite applies for the positive voltages. The high ER at low voltages is because of the fact that there is virtually no current for one polarization direction since the band tilting is not sufficient for FNT. The crossing bias for FNT and direct tunneling is basically the barrier height as expected. For thermionic injection the ER is quite large with its value given by the change in barrier height on polarization reversal due to its underlying mechanism. However, for a real device the change in total current is important which consists of all three contributions.

The total current density in the MFM structure for the two polarization states ($P > 0$ with solid lines and $P < 0$ with dashed lines) is shown in Fig. 3(a) for three different thicknesses $d$. It is basically governed by one of the three transport processes which might change with voltage, thickness, or polarization direction. The governing transport process can be read from Fig. 2. For a thickness up to 3.2nm and low voltages (voltages below the barrier height, here, approximately 1V) we obtain direct tunneling. This yields, for instance, an ER of about 40% at 3.2nm [Fig. 3(b)]. At a higher voltage FNT sets in with a decreasing and increasing ER at negative and positive *V*, respectively. A narrow transition region is in between both regimes, but apart from that region the current either due to FNT or direct tunneling can be neglected due to the steep increase in FNT with increasing voltage. For *V* > 0 it is possible to switch the major transport mechanism from FNT for *P* parallel to electron flow (*P* > 0) to direct tunneling for *P* antiparallel to electron flow (*P* < 0). This gives the highest ER (e.g. approximately 80% for 3.2nm, i.e., an "optimistic" ER of about 400%) in heterostructures with a thickness lower than about 3.2nm for reasonable voltages. At *V* < 0 this transition occurs when the FNT current branches cross each other yielding a drop in ER. For thicker films, for instance a 4.8nm ferroelectric film, thermionic injection is found for low voltages giving a high ER (positive at negative voltage and negative at positive voltage). At higher *V*, analogous to thinner films, FNT sets in yielding a transition region where we obtain either FNT or thermionic emission depending on the polarization direction.

Figure 4(a) shows the dependence of the ER on voltage *V* and thickness *d* of the ferroelectric layer in a contour plot. This map can be divided into four regions as sketched by the dashed lines. At low voltage and thickness we obtain direct tunneling with a transition to thermionic injection for thicker films. At high voltages FNT is the predominant transport mechanism with a transition to the other two mechanisms giving high ER and low ER at *V* > 0 and *V* < 0, respectively. The characteristic features of each transport mechanism and the transition regions on the map can be used to identify the transport mechanisms on similar maps for different parameters.

As already shown by Gruverman et al.[12] the ER increases with increasing thickness for direct tunneling [solid line in Fig. 4(b)]. We find the same for FNT [dashed-dotted line in Fig. 4(b)] as well as for thermionic injection (dashed line). However, the total current density decreases for thicker films by some orders of magnitude [see Fig. 3(a)]. It means that the high ER of thermionic injection is at the expense of the current density.

We tested the stability of the qualitative behavior of the transport mechanisms against reasonable changes in the used parameters, as exemplified in Fig. 5. Parameters related to a certain mechanism like $m_{e,ox}$ [shown in Fig. 5(a)] for direct tunneling and FNT, or $\varepsilon_{ifl}$ [Fig. 5(b)], $A^{**}$, and $T$ for thermionic-injection basically only shift the transition between tunneling and thermionic injection. For instance, a higher $m_{e,ox}$ shifts the transition to a lower thickness [Fig. 5(a)], whereas for higher $\varepsilon_{ifl}$ [see Fig. 5(b)], lower $A^{**}$, and lower $T$ the shift is to thicker ferroelectric films. In Figs. 5(c) and 5(d) parameters which are related to the change in potential on polarization reversal, namely, $P$ [in Fig. 5(c)] and $l_1$ and $l_2$ [in Fig. 5(d)] are changed. It is obvious from Eqs. (3)-(5) that a higher polarization, larger screening lengths, and lower permittivity $\varepsilon_{stat}$, which also directly influences the potential barrier, give higher ER, which gives the ideal structure. This is in agreement with Fig. 5(c), where the polarization $P$ is increased from 3 to $10\mu C/cm^2$. In Fig. 5(d) the screening capacitances of the two different electrodes are exchanged yielding a mirroring of the current densities with respect to the $y$ axis and an exchange of $P > 0$ with $P < 0$, giving in this case, for instance, a negative ER for direct tunneling.

The effect of changing the barrier height $\Phi_i$ at the interface between metal and ferroelectric $M_i/F$ is shown in Fig. 6. Thermionic injection is strongly dependent on the barrier height, with increasing current for decreasing barriers. Therefore, the shift of the transition from direct tunneling to thermionic injection to lower thicknesses for lower barriers [compare Figs. 4(a), 6(a), and 6(b) for decreasing barrier height] is reasonable. Furthermore, the transition from direct tunneling to FNT moves to lower voltage for lower barriers, i.e., the threshold for the onset of FNT is lowered.

For barriers, which are asymmetric even without polarization, i.e., $\Phi_1 \neq \Phi_2$, the resulting behavior regarding the onset of thermionic injection for $V > 0$ is similar to the one for $\Phi_1$, and the behavior for $V < 0$ is similar to $\Phi_2$ because those are the barriers which have to be overcome thermally. Hence, in Fig. 6(c), where $\Phi_1 = 1.4$eV and $\Phi_2 = 0.6$eV, thermionic injection is virtually absent for positive voltages, comparable to the case of $\Phi_1 = \Phi_2 = 1.4$eV and sets in at very low thickness for negative voltage like the case of $\Phi_1 = \Phi_2 = 0.6$eV. Interestingly, the transition between direct tunneling and FNT is at low positive voltages and at high negative voltage. This is because of the enhancement of the band bending due to different band offsets favoring FNT at positive voltages for the case of $\Phi_1 > \Phi_2$.

## IV. DISCUSSIONS

First, we analyze the interplay of polarization direction and screening ability of the electrodes with the transport mechanisms by analyzing Figs. 2, 4(a), and 5(d). For direct tunneling the current is lower when the polarization points to the metal electrode with the smaller screening length over permittivity $l_i/\varepsilon_M$ because this yields a higher average barrier as can be established from Eqs. (2), (4), and (5). Thermionic-injection currents are higher for polarization pointing against charge (electrons) flow, which is due to the lower potential barrier height $\Phi_{B,i}$ in this case. For FNT the situation is different. At low voltages (below the crossing voltage) the currents are higher for polarization parallel to charge flow, whereas it is the other way around at high voltage. At low applied voltage a sufficiently high band bending is crucial to render a triangular-shaped barrier necessary for FNT, which can be enhanced by the depolarization field. However, at high voltages when the potential barrier is sufficiently triangular for both polarization directions the barrier height $\Phi_{B,i}$, which hence determines the average barrier, makes the difference. Therefore the sign of ER, i.e., whether the current for $P > 0$ or $P < 0$ is higher, depends on the voltage polarity and on the major contributing transport mechanism, finally on thickness and voltage.

The major contributing mechanism at a certain voltage and thickness range depends also on the material properties of the MFM heterostructure, i.e., the parameters in our simulation. Therefore, we will discuss the applicability of the used parameters to real MFM heterostructures and their impact on transport and ER. As already mentioned in the previous section the parameters can be divided into two sets, some change the potential barrier and some are related only to a certain transport mechanism.

Since the potential barrier is the same for all transport mechanisms we start our discussion with $d$, $P$, $l_1/\varepsilon_{M,1}$, $l_2/\varepsilon_{M,2}$, $\Phi_1$, $\Phi_2$, and $\varepsilon_{stat}$. As shown in Fig. 4(b), a high thickness $d$ leads to a large ER for all transport mechanisms, but in turn it also results in a high resistance and correspondingly low current densities [see Fig. 3(a)] which might not be favorable for applications due to the long RC time constant related to it. Another way of obtaining high ER ratios is to employ thermionic injection which works for large voltage ranges, but it only predominates the transport at higher thickness with corresponding low conductivity. Furthermore, thermionic injection is strongly temperature dependent and this would alter the performance of a device used in an unstable temperature environment. A temperature independent, high ER at low thickness and therefore low resistivity can be achieved in the transition region between direct tunneling and FNT at positive voltage applied to the metal with the larger screening length over permittivity ratio, i.e., for instance V > 0 for $l_1/\varepsilon_{M,1} < l_2/\varepsilon_{M,2}$.

The bulk spontaneous polarization value $P_S$, e.g., about 30μC/cm$^2$ in BTO, might be higher than the one used in the simulations, but in ultrathin ferroelectric films $P$ decreases.[5] For instance, $P$ was found to be about 10μC/cm$^2$ in 5nm BTO (Ref. 2) and about 6μC/cm$^2$ in 4nm PbZr$_{0.20}$Ti$_{0.80}$O$_3$ (PZT).[1] We did not include the thickness dependence in our analysis. However, since the polarization increases with increasing thickness, this would add to the trend of increasing ER for increasing thickness [Fig. 4(b)]. Some polarization charges might also be screened by charges resulting from defects inside the ferroelectric. In this scenario polarization and screening charges are almost at the same place yielding no residual field and decreasing the polarization charge contributing to the depolarization field. This makes the used polarization value reasonable even for reported higher experimental values in ultrathin films.[32] Another mechanism yielding no residual field and therefore reducing the polarization charge contributing to our calculation is ionic screening. Since the depolarization field and the change in potential on polarization reversal is linear in $P$, higher ER is expected for a high unscreened $P$ value.

The static permittivity of the ferroelectric $\varepsilon_{stat}$ can be taken from capacitance-voltage measurements at ultrathin ferroelectric films. The preferred growth direction on standard substrates such as SrTiO$_3$ is usually $c$-axis oriented. Kim $et$ $al.$[33] report $\varepsilon_{stat}$ in the range of about 60-250 depending on the voltage for a 30nm BTO film with no significant dependence on thickness down to 5nm. The value of $\varepsilon_{stat}$ is quite close to the bulk value of 160 for *c*-axis oriented single crystals.[34] We use the high-field permittivity 60 in our simulations since this is the value free from extrinsic contributions, i.e., domain walls or alignment of defect dipoles. A lower $\varepsilon_{stat}$ of the ferroelectric, as, e.g., found in PZT (Ref. 35) or $PbTiO_3$,[36] is advantageous for larger effects as can be seen from Eqs. (3) and (4).

Usually, high-quality oxide ferroelectric films are grown epitaxially onto a perovskite electrode, e.g., SRO, whereas the top electrode might be a suitable elemental metal. Good ferroelectric properties were shown, for instance, for Cu on PZT.[37] Therefore, the screening lengths $l_i$ reported in literature, calculated from the free-electron model, for Cu and SRO (0.55Å and 0.8Å, respectively)[24,23] were used as well as the corresponding permittivties $\varepsilon_M$, which together yield the capacitance of the screening charge. Other metals with corresponding different capacitance of the screening space charge would change the magnitude of the ER. The ER is enhanced either for bad metals with large screening lengths $l_i/\varepsilon_{M,i}$ and hence a small capacitance of the screening charges at both interfaces [compare Fig. 7(a) to Fig. 4(a)] or even more for highly different screening lengths $l_i/\varepsilon_{M,i}$ at top and bottom interfaces as shown in Fig. 7(b). This applies for all three transport mechanisms, since large (or a large difference in) screening lengths $l_i/\varepsilon_{M,i}$ increase the change in barrier offset and depolarization field, which is important for FNT and thermionic injection, as well as the difference in average barrier height important for direct tunneling. Of course, if the screening is too low and the polarization too high, the depolarization field might suppress ferroelectricity completely[26] or favor formation of a pinned 180° domain state.[38] The choice of metal may also influence the barrier height $\Phi_i$.

Interestingly, we find an ER even for a symmetric MFM structure ($M_1 = M_2$). Figure 7 (c) shows that at low voltage and thickness, where direct tunneling dominates the transport, the ER is vanishing in agreement with Ref. [8]. At higher voltage and thickness the ER is due to FNT and thermionic injection, respectively, like in the asymmetric case. The ER for FNT and thermionic injection is evident because both mechanisms strongly depend on the barrier properties at the electron injecting electrode. Indeed, these properties, i.e., the barrier height and the depolarization field, change on polarization reversal as sketched in Fig. 1(b) and therefore yield a rather high ER.

The ER due to switching between direct tunneling and FNT peaks at about 1V for $\Phi_i$ = 1V [see Fig. 3(b)]. This gives a rather high field and might be close to the coercive field of the ferroelectric. At the coercive field the polarization direction is switched if the electrical field is antiparallel to the polarization, which in turn means that a polarization-induced electro-resistance effect above the coercive field is not accessible by any experiment. But, as shown in Fig. 6(a), the peak in ER can be shifted to a lower voltage by decreasing the barrier height with the shift being approximately proportional to the barrier height. As shown in Fig. 6(c) it is possible to combine an epitaxial bottom electrode (such as SRO, which gives a high theoretical band offset of $\Phi_i$ = 1.8V on BTO) (Ref. 39) with a metal giving a low band offset suppressing thermionic injection at the voltage polarity which is favorable for FNT. At least, in principle, the potential barrier height between ferroelectric and top electrode can be determined by the metal work function, i.e., it can be selected by changing the top electrode material.[15, 18, 37] But it also depends on interface states.[15] The useful band offsets might also be limited if the difference between the two metals is too high since the resulting band bending might generate backswitching. This means it favors one polarization direction, whereas the other is unstable.

These parameters change the ER directly, but the choice in $P$ and $\varepsilon_{\text{stat}}$ is limited by the ferroelectric material. Therefore, the most versatile playground for optimizing the ER is the capacitance of the screening charges, i.e., the screening length and ionic permittivity, and the band offset of the metal electrodes. Especially in real devices the top electrode offers the highest flexibility, being not restricted by the growth of the ferroelectric layer.

The parameters related to specific transport processes are also material properties of the ferroelectric and therefore cannot be widely changed. The effective tunneling mass $m_{e,\text{ox}}$ is specific for the two quantum-mechanical mechanisms, namely, direct tunneling and FNT. In our simulations we use the free-electron mass $m_e$ as effective tunneling mass $m_{e,\text{ox}}$ which is given in literature in the range of $1 m_e$ (Ref. 12) to $5 m_e$ (Ref. 10) for BTO. A higher effective tunneling mass might be expected, although the effective tunneling mass is not necessarily equal to the effective mass in the bulk ferroelectric. However, a higher $m_{e,\text{ox}}$ would just suppress tunneling which in turn leads to a transition to thermionic injection at lower film thickness [Fig. 5(a)].

The parameters connected with thermionic injection are Richardson's constant $A^{**}$ and $\varepsilon_{\text{ifl}}$. For the effective $A^{**}$ only few reports with strongly varying values exist, therefore a standard value for semiconductors is used. The permittivity in thermionic injection is the image force lowering permittivity $\varepsilon_{\text{ifl}}$. This is not necessarily the static or the optic one.[15] Using the approach by Sze,[15] one arrives at the terahertz range for a typical ferroelectric, where the permittivity might be a bit higher than in the optical range.[40] This permittivity $\varepsilon_{\text{ifl}} = 10$ is also sufficiently high to get only small changes in the potential barrier shape.

However, as already qualitatively shown (see Fig. 5), these parameters primarily change the thickness and voltage range where the transition between the transport mechanisms occurs.

## V. CONCLUSIONS

We studied the transport mechanisms and electro-resistive effects in ferroelectric tunnel junctions with a thickness up to 5nm including direct tunneling, FNT, and thermionic-injection currents. All three mechanisms concurrently govern the transport with the major contribution depending on thickness, voltage, and polarization direction, as well as several parameters linked to the materials used in the MFM. The sign of the ER depends on voltage polarity and on the major mechanism contributing to the current. All transport mechanisms yield higher ER ratios at higher ferroelectric film thickness but the device performance may suffer from the high resistance at high film thickness. Most versatile for optimizing the ER are the metal electrodes. The metal electrodes govern the value of ER by their screening ability and additionally influence the potential barrier height between ferroelectric and electrode, which controls the voltage range where a transition between different transport mechanisms occurs. Temperature independent, high ER combined with low resistivity at an acceptable voltage can be obtained by employing the change in transport mechanism by polarization switching from direct tunneling to FNT. The results are qualitatively stable against reasonable changes in parameters with highest effects for large $P$, small $\varepsilon_\text{stat}$, small potential barriers, and large $l_\text{i}/\varepsilon_\text{M,I}$, and should qualitatively apply for all tetragonal ferroelectrics sandwiched between metal electrodes.

**ACKNOWLEDGEMENTS**

The authors would like to thank Dietrich Hesse for carefully reading this manuscript. This work has been supported by the German Science Foundation (DFG) through SFB762.

**Figure captions**

FIG. 1. (Color online) (a) Sketch of the MFM heterostructure with $P > 0$ and (b) the corresponding energy potential profile of the potential barrier for $P > 0$ (solid line) and $P < 0$ (dashed line), respectively.

FIG. 2. (Color online) Voltage dependence of the current density contributions $j$ from DT, FNT, and TI of (a) 1.2nm, (b) 3.2nm, and (c) 4.8nm thick ferroelectric to the total current for $P > 0$ (solid lines) and $P < 0$ (dashed lines), respectively. The following parameters are used in the simulation: $P = 3\mu C/cm^2$, $\varepsilon_{stat} = 60$, $\varepsilon_{ifl} = 10$, $m_{e,ox} = m_e$, $A^{**} = 10^6 Am^{-2}K^{-2}$, $l_1 = 0.8$Å, $l_2 = 0.55$Å, $\varepsilon_{M,1} = 8$, $\varepsilon_{M,2} = 2$, and $\Phi_1 = \Phi_2 = 1V$.

FIG. 3. (Color online) Voltage dependence of the (a) total current density $j_{tot}$ for $P > 0$ (solid lines) and $P < 0$ (dashed lines), respectively, and (b) the corresponding ER for a 1.2, 3.2, and 4.8 nm thick ferroelectric. The parameters are the same as in Fig. 2.

FIG. 4. (Color online) (a) Contour plot of the *ER* versus voltage *V* and ferroelectric film thickness $d$ with following parameters: $P = 3\mu C/cm^2$, $\varepsilon_{stat} = 60$, $\varepsilon_{ifl} = 10$, $m_{e,ox} = m_e$, $A^{**} = 10^6 Am^{-2}K^{-2}$, $l_1 = 0.8$Å, $l_2 = 0.55$Å, $\varepsilon_{M,1} = 8$, $\varepsilon_{M,2} = 2$, and $\Phi_1 = \Phi_2 = 1V$. The spacing between the thin solid lines represents a change in ER of 0.04. The transition regions between DT, FNT, and TI are sketched by the thick dashed lines; (b) a cross section

through the contour plot at $V = 1.5$V (dashed-dotted line) and $V = -0.2$V (solid and dashed line), i.e., the thickness dependence of the ER.

FIG. 5. (Color online) ER (contour plot) versus applied voltage $V$ and film thickness $d$ for different parameters (a) $m_{e,ox} = 3m_e$, (b) $\varepsilon_{ifl} = 60$, (c) $P = 10\mu C/cm^2$, and (d) $l_1 = 0.55$Å, $l_2 = 0.8$Å, $\varepsilon_{M,1} = 2$, and $\varepsilon_{M,2} = 8$. The scale and the other parameters are the same as in Fig. 4.

FIG. 6. (Color online) ER (contour plot) versus applied voltage $V$ and film thickness $d$ showing the dependence on potential barrier $\Phi$ with (a) $\Phi_1 = \Phi_2 = 0.6$V, (b) $\Phi_1 = \Phi_2 = 1.4$V, and (c) $\Phi_1 = 1.4$V and $\Phi_2 = 0.6$V. The scale and the other parameters are the same as in Fig. 4.

FIG. 7. (Color online) ER (contour plot) versus applied voltage $V$ and film thickness $d$ showing the dependence on screening lengths $l_1$ and $l_2$ with (a) $l_1 = 3.2$ Å and $l_2 = 2.2$Å, (b) $l_1 = 0.8$Å and $l_2 = 2.2$Å, and (c) $l_1 = l_2 = 0.8$Å and $\varepsilon_{M,1} = \varepsilon_{M,2} = 8$. The scale and the other parameters are the same as in Fig. 4.

[1] V. Nagarajan, S. Prasertchoung, T. Zhao, H. Zheng, J. Ouyang, R. Ramesh, W. Tian, X. Q. Pan, D. M. Kim, C. B. Eom, H. Kohlstedt, and R. Waser, Appl. Phys. Lett. **84**, 5225 (2004).

[2] Y. S. Kim, D. H. Kim, J. D. Kim, Y. J. Chang, T. W. Noh, J. H. Kong, K. Char, Y. D. Park, S. D. Bu, J.-G. Yoon, and J.-S. Chung, Appl. Phys. Lett. **86**, 102907 (2005).

[3] S. K. Streiffer, J. A. Eastman, D. D. Fong, C. Thompson, A. Munkholm, M. V. Ramana Murty, O. Auciello, G. R. Bai, and G. B. Stephenson, Phys. Rev. Lett. **89**, 067601 (2002).

[4] I. Vrejoiu, G. Le Rhun, L. Pintilie, D. Hesse, M. Alexe, and U. Gösele, Adv. Mater. **18**, 1657 (2006).

[5] J. Junquera and P. Ghosez, Nature **422**, 506 (2003).

[6] E. Y. Tsymbal and H. Kohlstedt, Science **313**, 181 (2006).

[7] H. Kohlstedt, N. A. Pertsev, J. Rodriguez Contreras, and R. Waser, Phys. Rev. B **72**, 125341 (2005).

[8] M. Y. Zhuravlev, R. F. Sabirianov, S. S. Jaswal, and E. Y. Tsymbal, Phys. Rev. Lett. **94**, 246802 (2005).

[9] M. Gajek, M. Bibes, S. Fusil, K. Bouzehouane, J. Fontcuberta, A. Barthélémy, and A. Fert, Nature Mater. **6**, 296 (2007).

[10] V. Garcia, M. Bibes, L. Bocher, S. Valencia, F. Kronast, A. Crassous, X Moya, S. Enouz-Vedrenne, A. Gloter, D. Imhoff, C. Deranlot, N. D. Mathur, S. Fusil, K. Bouzehouane, and A. Barthélémy, Science **327**, 1106 (2010).

[11] V. Garcia, S. Fusil, K. Bouzehouane, S. Enouz-Vedrenne, N. D. Mathur, A. Barthélémy, and M. Bibes, Nature (London) **460**, 81 (2009).

[12] A. Gruverman, D. Wu, H. Lu, Y. Wang, H. W. Jang, C. M. Folkman, M .Y. Zhuravlev, D. Felker, M. Rzchowski, C. B. Eom, and E. Y. Tsymbal, Nano Lett. **9**, 3539 (2009).

[13] P. Maksymovych, S. Jesse, P. Yu, R. Ramesh, A. P. Baddorf, and S. V. Kalinin, Science **324**, 1421 (2009).

[14] L. Pintilie and M. Alexe, J. Appl. Phys. **98**, 124103 (2005).

[15] S.M. Sze, *Physics of Semiconductor Devices*, 3rd ed. (Wiley-Interscience, , Hoboken, NJ, 2007).

[16] J. Shin, V. B. Nascimento, A. Y. Borisevich, E. W. Plummer, S. V. Kalinin, and A. P. Baddorf, Phys. Rev. B **77**, 245437 (2008).

[17] A. Crassous, V. Garcia, K. Bouzehouane, S. Fusil, A. H. G. Vlooswijk, G. Rispens, B. Noheda, M. Bibes, and A. Barthélémy, Appl. Phys. Lett. **96**, 042901 (2010).

[18] J. Robertson and C. W. Chen, Appl. Phys. Lett. **74**, 1168 (1999).

[19] M. Dawber, K. Rabe, and J. F. Scott, Rev. Mod. Phys. **77**, 1083 (2005).

[20] R. R. Mehta, B. D. Silverman, and J.T. Jacobs, J. Appl. Phys. **44**, 3379 (1973).

[21] N. Pertsev and H. Kohlstedt, Phys. Rev. Lett. **98**, 257603 (2007).

[22] G. Gerra, A. Tagantsev, N. Setter, and K. Parlinski, Phys. Rev. Lett. **96**, 107603 (2006).

[23] D. Kim, J. Jo, Y. Kim, Y. Chang, J. Lee, J.-G. Yoon, T. Song, and T. Noh, Physi. Rev. Lett. **95**, 237603 (2005).

[24] H. Ibach and H. Lüth, *Solid State Physics*, 4th ed. (Springer, New York, 2009), p. 152.

[25] N.W. Ashcroft and N.D. Mermin, *Solid State Physics* (Brooks-Cole, Belmont, MA,1976), p. 340.

[26] P. Ghosez and J. Junquera, in *Handbook of Theoretical and Computational Nanotechnology*, edited by M. Rieth and W. Schommers (American Scientific Publisher, Stevenson Ranch, CA, 2006).

[27] D. Fong, C. Cionca, Y. Yacoby, G. Stephenson, J. Eastman, P. Fuoss, S. Streiffer, C. Thompson, R. Clarke, R. Pindak, and E. Stern, Physi. Rev. B **71**, 144112 (2005).

[28] J. G. Simmons, J. Appl. Phys. **34**, 1793 (1963).

[29] R. H. Fowler and L. Nordheim, Proc. R. Soc. London, Ser.A **119**, 173 (1928).

[30] Please note that the polarization entering the calculation is not the spontaneous polarization but the polarization which is screened by Thomas-Fermi screening.

[31] Reports on the permittivity of Cu are rare. Since it is not ionic we assume a smaller permittivity than the one of SRO. $\varepsilon_M = 2$ is in agreement with Ref. 23.

[32] A. Petraru, H. Kohlstedt, U. Poppe, R. Waser, A. Solbach, U. Klemradt, J. Schubert, W. Zander, and N. A. Pertsev, Appl. Phys. Lett. **93**, 072902 (2008).

[33] Y. S. Kim, J. Y. Jo, D. J. Kim, Y. J. Chang, J. H. Lee, T. W. Noh, T. K. Song, J.-G. Yoon, J.-S. Chung, S. I. Baik, Y.-W. Kim, and C. U. Jung, Appl. Phys. Lett. **88**, 072909 (2006).

[34] R. Landauer, D. R. Young, and M. E. Drougard, J. Appl. Phys. **27**, 752 (1956).

[35] L. Pintilie, I. Vrejoiu, D. Hesse, G. Le Rhun, and M. Alexe, Phys. Rev. B **75**, 224113 (2007).

[36] J.A. Sanjurjo, E. Lopez-Cruz, and G. Burns, Phys. Rev. B **28**, 7260 (1983).

[37] L. Pintilie, I. Vrejoiu, D. Hesse, and M. Alexe, J. Appl. Phys. **104**, 114101 (2008).

[38] V. Nagarajan, J. Junquera, J. Q. He, C. L. Jia, R. Waser, K. Lee, Y. K. Kim, S. Baik, T. Zhao, R. Ramesh, P. Ghosez, and K. M. Rabe, J. Appl. Phys. **100**, 051609 (2006).

[39] J. Junquera, M. Zimmer, P. Ordejón, and P. Ghosez, Phys. Rev. B **67**, 155327 (2003).

[40] T. Tsurumi, J. Li, T. Hoshina, and H. Kakemoto, Appl. Phys. Lett. **91**, 182905 (2007).

FIG. 1:

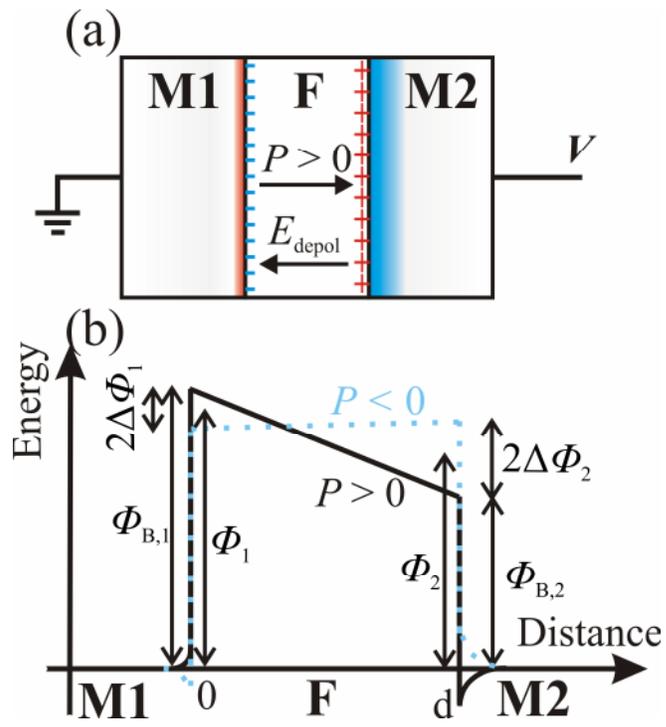

FIG. 2:

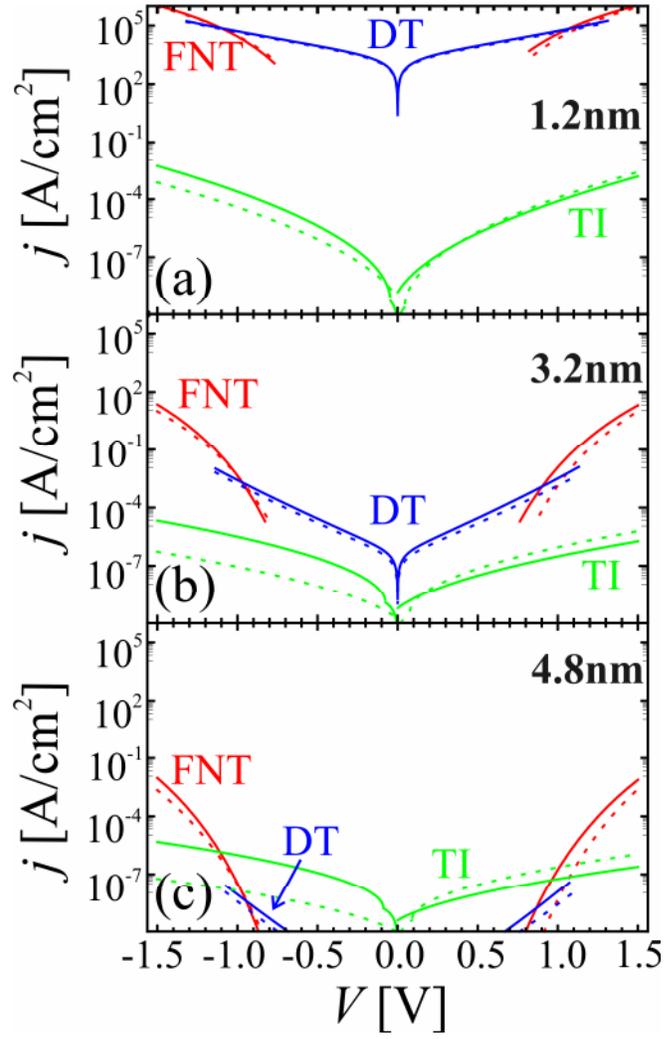

FIG. 3:

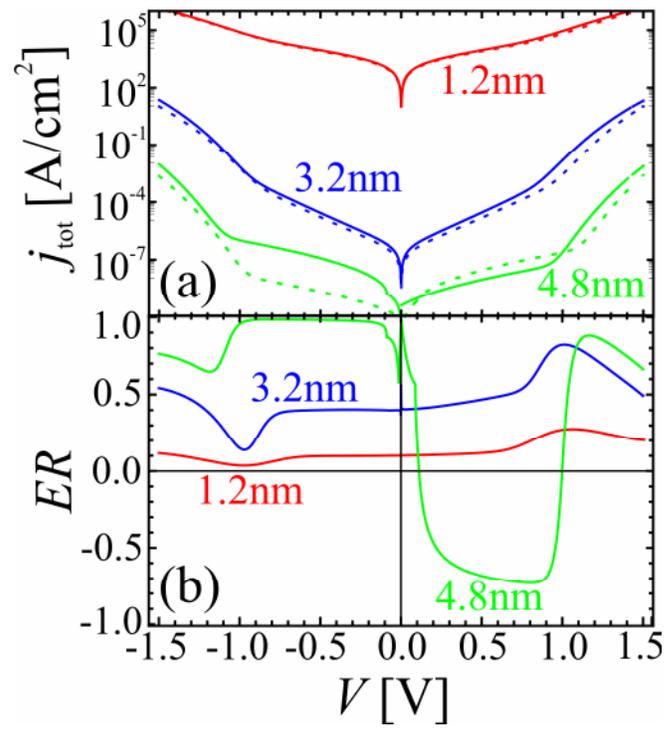

FIG. 4:

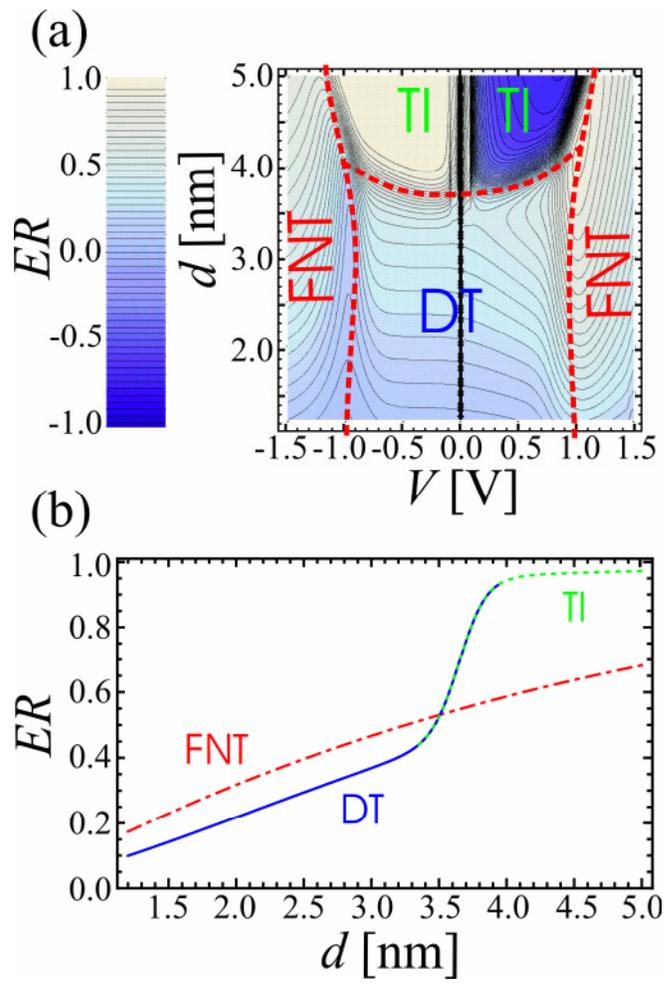

FIG. 5:

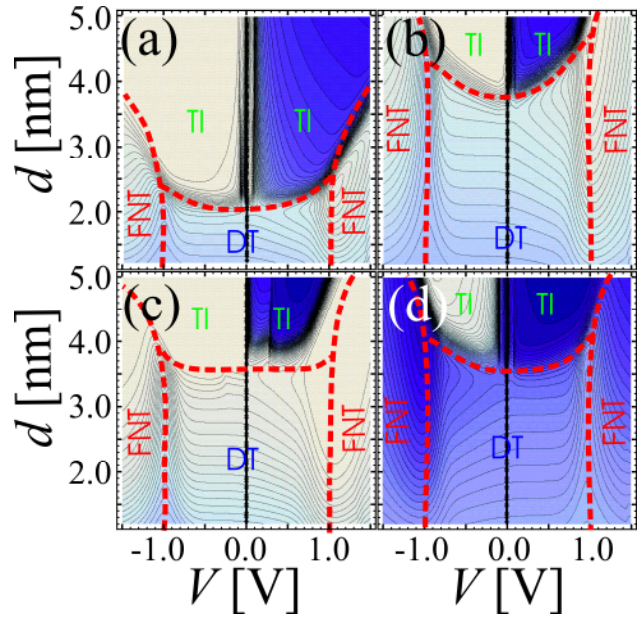

FIG. 6:

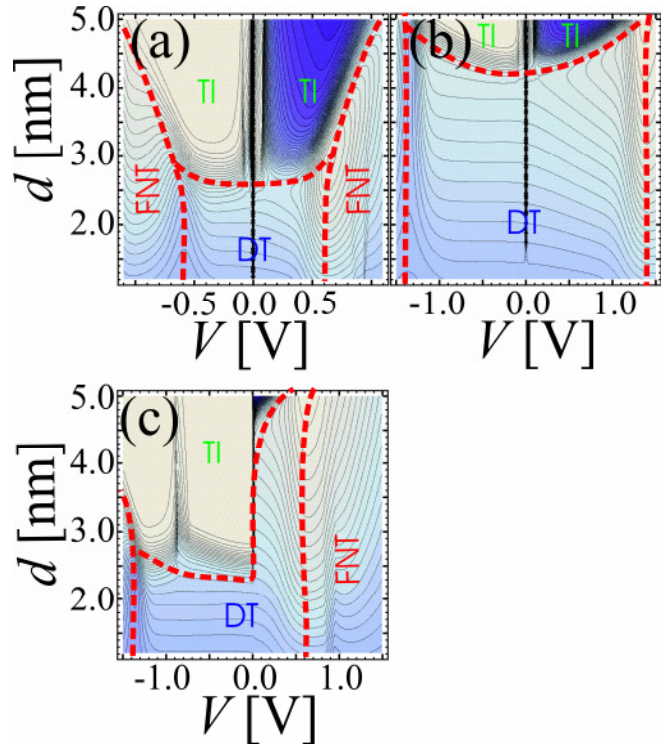

FIG. 7:

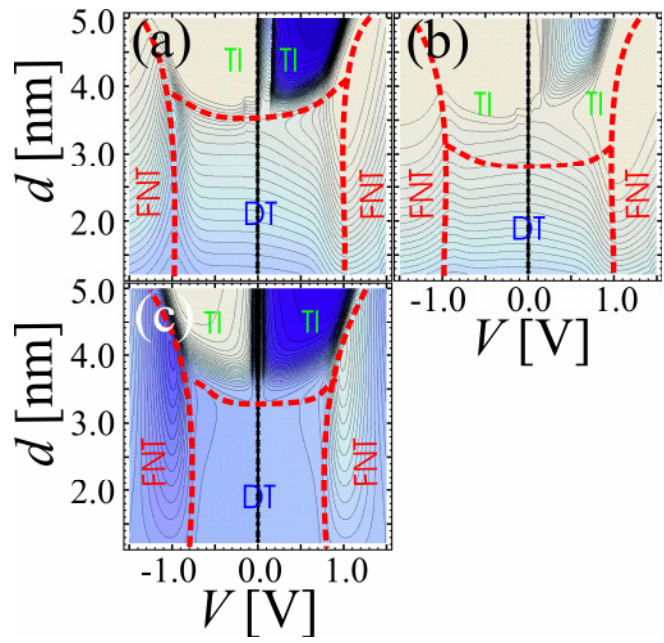